\documentclass[journal]{IEEEtran}
\ifCLASSINFOpdf
\usepackage[pdftex]{graphicx}
\graphicspath{figures/}
\DeclareGraphicsExtensions{.pdf,.jpeg,.png}
\else
\usepackage[dvips]{graphicx}
\graphicspath{{figures/}}
\DeclareGraphicsExtensions{.eps}
\fi
\usepackage[cmex10]{amsmath}
\usepackage{bm}
\usepackage{hyperref}
\usepackage{mathtools}
\usepackage{cite}
\usepackage{amsfonts}
\usepackage{url}
\begin{document}
\title{Exploiting Oxide Based Resistive RAM Variability for Bayesian Neural Network Hardware Design}

\author{Akul~Malhotra, Sen~Lu, Kezhou~Yang, 
and~Abhronil~Sengupta,~\IEEEmembership{Member,~IEEE}
\thanks{Manuscript received January, 2020.}
\thanks{The authors are with the School
of Electrical Engineering and Computer Science, Department of Materials Science and Engineering, The Pennsylvania State University, University Park,
PA 16802, USA. A. Malhotra is also affiliated with Birla Institute of Technology and Science, Pilani, Rajasthan 333031, India. E-mail: sengupta@psu.edu.}}
\maketitle
\begin{abstract}
\small{Uncertainty plays a key role in real-time machine learning. As a significant shift from standard deep networks, which does not consider any uncertainty formulation during its training or inference, Bayesian deep networks are being currently investigated where the network is envisaged as an ensemble of plausible models learnt by the Bayes' formulation in response to uncertainties in sensory data. Bayesian deep networks consider each synaptic weight as a sample drawn from a probability distribution with learnt mean and variance. This paper elaborates on a hardware design that exploits cycle-to-cycle variability of oxide based Resistive Random Access Memories (RRAMs) as a means to realize such a probabilistic sampling function, instead of viewing it as a disadvantage. }
\end{abstract}


\section{Introduction}
While Bayesian deep learning has shown promise to serve as a pathway for enabling Probabilistic Machine Learning, the algorithms have been primarily developed without any insights regarding the underlying hardware implementation. Bayesian techniques are more computationally expensive than their non-Bayesian counterparts, thereby limiting their training and deployment in resource-constrained environments like wearables and mobile edge devices. In addition to the standard von-Neumann bottleneck \cite{zidan2018future} prevalent in current deep learning networks (where memory access and memory leakage can account for significant portion of the total energy consumption profile), Bayesian deep learning involves repeated sampling of network weights from learnt probability distributions (in most cases, Gaussian distributions are used which are much more hardware expensive than uniform probability distributions) and inference based on the sampled weights. For instance, implementing just a single synapse would involve a costly CMOS Gaussian random number generator circuit. Repeated parameter sampling and evaluation for just a single inference operation worsens the von-Neumann bottleneck issue further. With deep networks involving more than a million parameters coupled with the necessity of performing mathematical operations on sampled probabilistic data, the projected hardware costs for a typical CMOS implementation would be humongous. Thus there is a dire need to rethink hardware designs for such probabilistic machine learning frameworks ground-up where the core hardware units are better matched to the models of computation.
Our paper elaborates on a cohesive design of an RRAM-based Bayesian processor that leverages benefits of cycle-to-cycle variability for gaussian random number sampling and ``In-Memory" crossbar architectures to realize energy efficient hardware primitives that have the potential of enabling probabilistic Artificial Intelligence. 
Non-idealities and stochasticity prevalent in RRAM technologies have been typically not exploited for computing \cite{dalgaty2019hybrid}. While there have been recent efforts at incorporating various aspects of RRAM stochastic switching processes for neuromorphic \cite{suri2015neuromorphic,ly2018role,suri2013bio} and hardware security \cite{yu2017rram,wei2016true,chen2015comprehensive,sahay2017oxram,pang2017optimization} applications, the concept of utilizing probability distributions obtained from RRAM cycle-to-cycle variations for Bayesian deep learning applications have been not explored before. 

\section{Utilizing Cycle-to-Cycle RRAM Variability for Probability Distribution Sampling}
\begin{figure}
\centering
\includegraphics[width=0.48\textwidth]{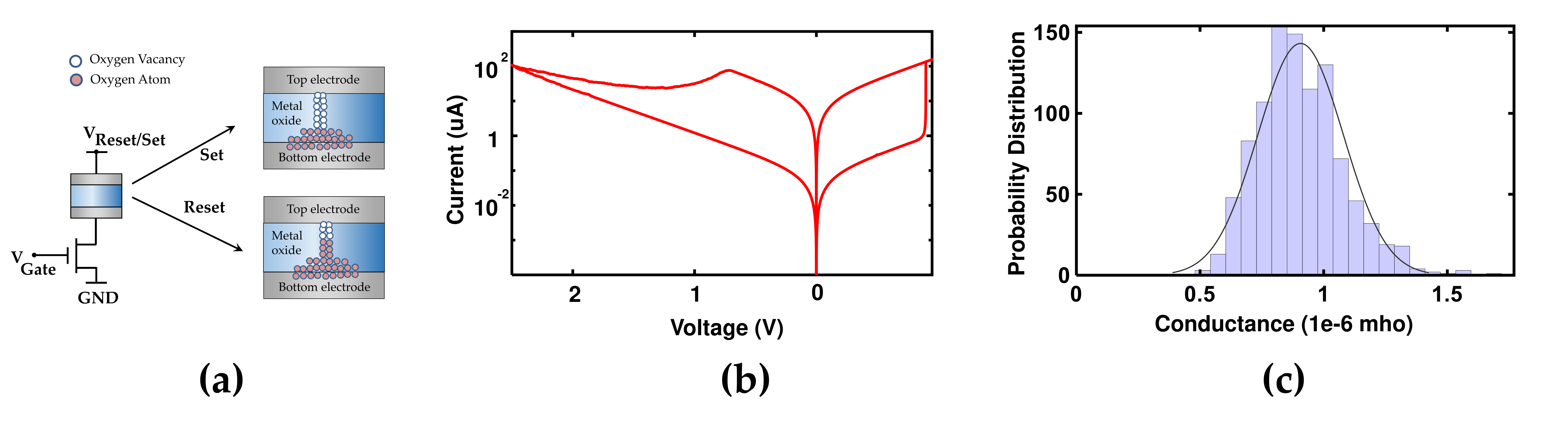}
\caption{(a) A 1T-1R structure where the magnitude and sign of $V_{Reset/Set}$ controls the device resistance. (b) $I-V$ characteristics of our RRAM model based on parameters reported in Ref. \cite{jiang2016compact} for a bilayer $TiO_{x}/HfO_{x}$ material stack with $3.3nm$ thick $HfO_{x}$ layer. }
\label{fig1}
\vspace{-2mm}
\end{figure}
Metal oxide RRAMs are emerging as an alternative to traditional CMOS based technology in a plethora of Boolean and non-Boolean computing applications due to their high density, CMOS compatibility and low power consumption, among others. A metal oxide RRAM is a two-terminal device whose resistance can be changed upon applying a voltage across the terminals (see Fig. \ref{fig1}). The device consists of an oxide layer sandwiched between two electrode layers. 

The metal oxide RRAM has two extreme resistance states, the high resistance state (HRS) and the low resistance state (LRS). When the resistance of the RRAM switches from HRS to LRS, the event is called a SET process and when the resistance switches from LRS to HRS, the event is called a RESET process. The SET and RESET processes can be explained by the growth and rupture of a  conduction filament in the oxide. The distance between the tip of the filament and the opposite electrode, or the gap length ($g$), is the primary variable governing the device $I-V$ characteristics. During the SET process, the oxygen ions drift to the anode interface, creating a conductive oxygen vacancy path which leads to the growth of the conduction filament, reducing the gap length. During the RESET process, the reverse electric field drives the oxygen ions back to recombine with the vacancies, breaking down the conduction filament and increasing the gap length. The current flowing through the device is generalized using the following equation \cite{jiang2016compact},
\begin{equation}
    I=I_o.\exp{(-\frac{g}{g_o})}.\sinh{(\frac{V}{V_o})}
\end{equation}
where, $I_o$, $G_o$ and $V_o$ are fitting parameters, $g$ is the gap length and $V$ is the voltage across the device. The exponential dependence on the gap length arises due to the various tunneling mechanisms present in the device \cite{jiang2016compact}. 
We used an experimentally calibrated publicly accessible RRAM model \cite{nanoHUB.org19} and our simulation parameters correspond to a bilayer $TiO_{x}/HfO_{x}$ material stack with $3.3nm$ thick $HfO_{x}$ layer.  Please refer to Ref. \cite{jiang2016compact} for a description of the device model and fitting parameters.
\begin{table}[h]
\label{table2}
\center
\centerline{TABLE I. RRAM Device Simulation Parameters}
\vspace{2mm}
\begin{tabular}{c c}
\hline \hline
\bfseries Parameters & \bfseries Value\\
\hline
$g_0$ & $0.15 nm$ \\
$V_0$ & $ 0.35 V$ \\
$I_0$ & $ 0.2 mA$ \\
$v_0$ & $ 5\times 10^6 m/s$ \\
$\gamma_0$ & $20.8$ (SET), 24.9 (RESET) \\
$\beta_0$ & $4.8$ (SET), $19.0$ (RESET) \\
$\alpha$ & $1$\\
\hline \hline
\end{tabular}\\ 
\end{table}
\begin{figure}
\centering
\includegraphics[width=0.3\textwidth]{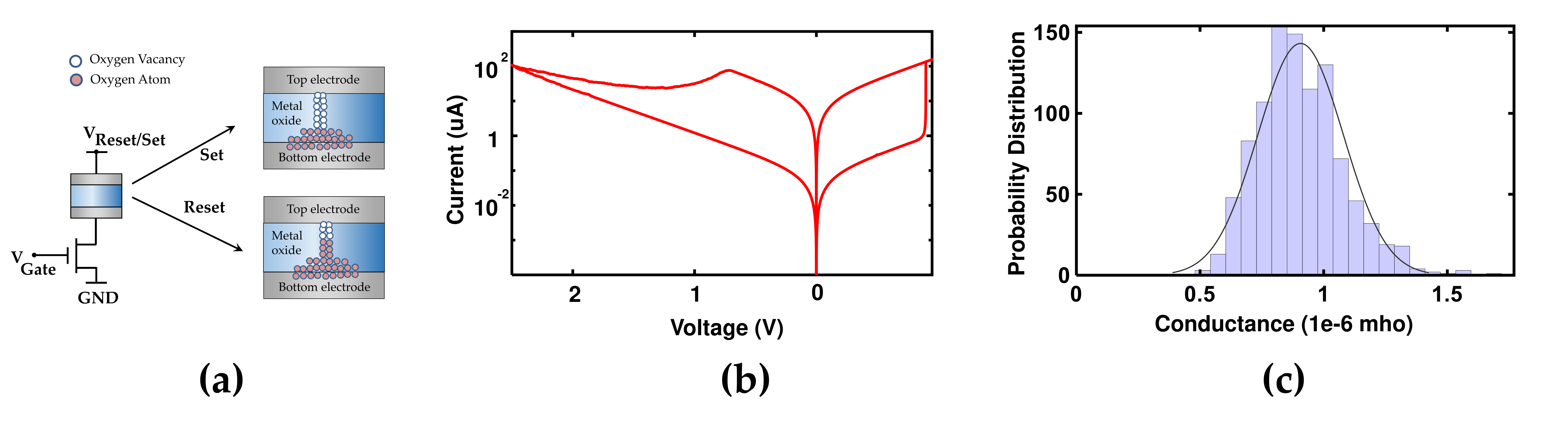}
\caption{Probability distribution of the cycle-to-cycle variation of the HRS resistance. The device is first SET using 1.1V and then RESET using -3V. Each cycle is $30ns$ in duration. The histogram represents the sampled RRAM conductance value at the end of the $30ns$ read phase. 1,000 datapoints have been used to plot the probability distribution.}
\label{fig2}
\vspace{-2mm}
\end{figure}

The RRAM RESET process is a gradual one and exhibits cycle-to-cycle resistance variation as depicted in Fig. \ref{fig2}. This is due to the random oxygen-vacancy filament dissolution process. While prior works have reported similar variation studies, they have typically considered this as a non-ideality. In this article, we utilize such a probability distribution sampling function for Bayesian network implementation. It is worth noting here that the distribution is not strictly Gaussian (log-normal is a better fit \cite{dalgaty2019hybrid,garbin2014variability}). However, we did not observe the minor deviation from the Gaussian distribution to impact the network accuracy. The network can be also inherently trained assuming log-normal probability priors to account for the device constraints. In this work, we have assumed Gaussian probability priors for training.

\section{System Design and Results} 

\begin{figure}[b!]
\vspace{-4mm}
\centering
\includegraphics[width=0.48\textwidth]{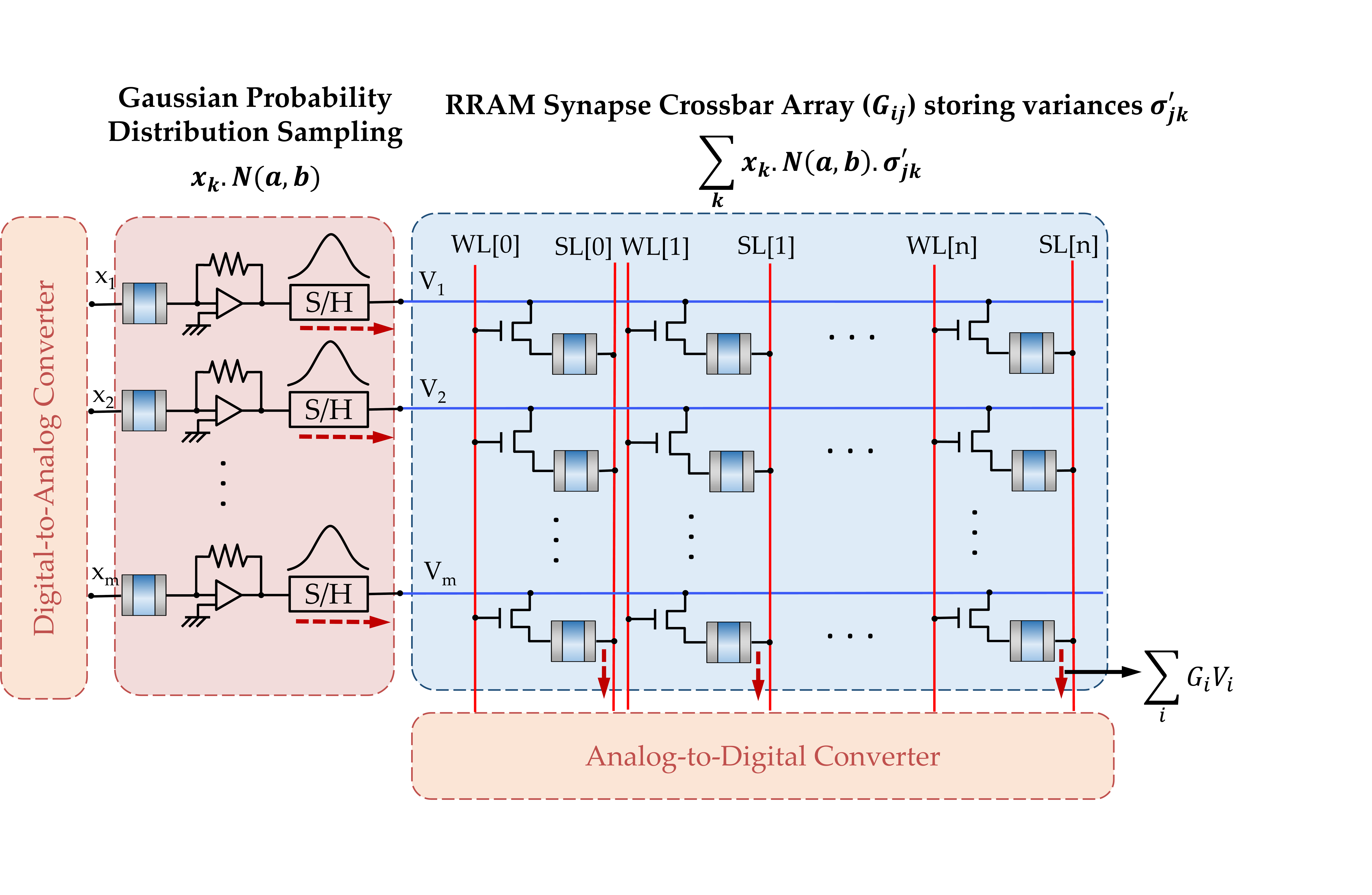}
\caption{RRAM Based Bayesian Neural Network Implementation. The rows of the crossbar array are driven by scaled inputs where the scaling factor is a sampling from a Normal distribution. The sampling operation can be simply implemented by an RRAM device interfaced with an amplifier and Sample-Hold circuit.}
\label{fig3}
\end{figure}
In a Bayesian Neural Network, the synaptic weights, $\textbf{W}$, are typically characterized by Gaussian probability distributions. Once all the \textit{posterior} distributions are learnt ($\mu$ and $\sigma$ parameters of the weight distributions) \cite{gal2015bayesian}, the network output corresponding to input, $\textbf{x}$, should be obtained by averaging the outputs obtained by sampling weights from the \textit{posterior} distribution of the weights, $\textbf{W}$ \cite{cai2018vibnn}. The output of the network, $y$, is therefore given by,
\begin{equation}
    y = \mathbb{E}_{P(\textbf{W}|D)}[f(\textbf{x,W})] \approx \mathbb{E}_{q(\textbf{W},\theta)}[f(\textbf{x,W})] \approx \frac{1}{S}\sum_{i=1}^{S}f(\textbf{x,W}^i)
\end{equation}
where, $P(D|\textbf{W})$ is the \textit{likelihood}, corresponding to the feedforward pass of the network, $f(\textbf{x,W})$ is the network mapping for input $\textbf{x}$ and weights, $\textbf{W}$. The approximation is performed over $S$ independent Monte Carlo samples drawn from the Gaussian distribution, $q(\textbf{W, $\theta$})$, characterized by parameters, $\theta=(\mu,\sigma$), where $\mu$ and $\sigma$ represent the mean and standard deviation vectors for the probability distributions representing $P(\textbf{W}|D)$ \cite{ghahramani2001propagation}. We consider the Variational Inference method \cite{houthooft2016vime,jordan1999introduction} for training the network in this work. 

Hence, the core computation in a Bayesian network is a dot-product between an input vector, $\textbf{x}$, and samples drawn from a Gaussian probability distribution, \textbf{$N$(\boldsymbol{$\mu$}, \boldsymbol{$\sigma$})}. Considering just a single layer and neglecting the neural transfer function, $f(\textbf{x,W}^i)$ for the $j$-th neuron can be simplified as,
\begin{equation}
   f(\textbf{x,W}^i_j) = \sum_{k}x_k . N(\mu_{jk},\sigma_{jk}) = \sum_{k}x_k . (\mu_{jk}^{\prime} + \sigma_{jk}^{\prime}.N(a,b)) 
\end{equation}
where, $k$ is the dimensionality of the input $\textbf{x}$, $N(\mu_{jk},\sigma_{jk})$ represent a particular weight sample drawn from a Normal probability distribution with mean $\mu_{jk}$ and variance $\sigma_{jk}$ and $N(a,b)$ is the probability distribution of cycle-to-cycle resistance variation with mean $a$ and variance $b$ obtained from the RRAM device utilized for probabilistic sampling (Fig. \ref{fig2}). It can be proved by simple algebraic manipulation that, $\mu_{jk}^{\prime}=\mu_{jk}-a/b$ and $\sigma_{jk}^{\prime}=\sigma_{jk}/b$.
\begin{figure}[t]
\vspace{-4mm}
\centering
\includegraphics[width=0.48\textwidth]{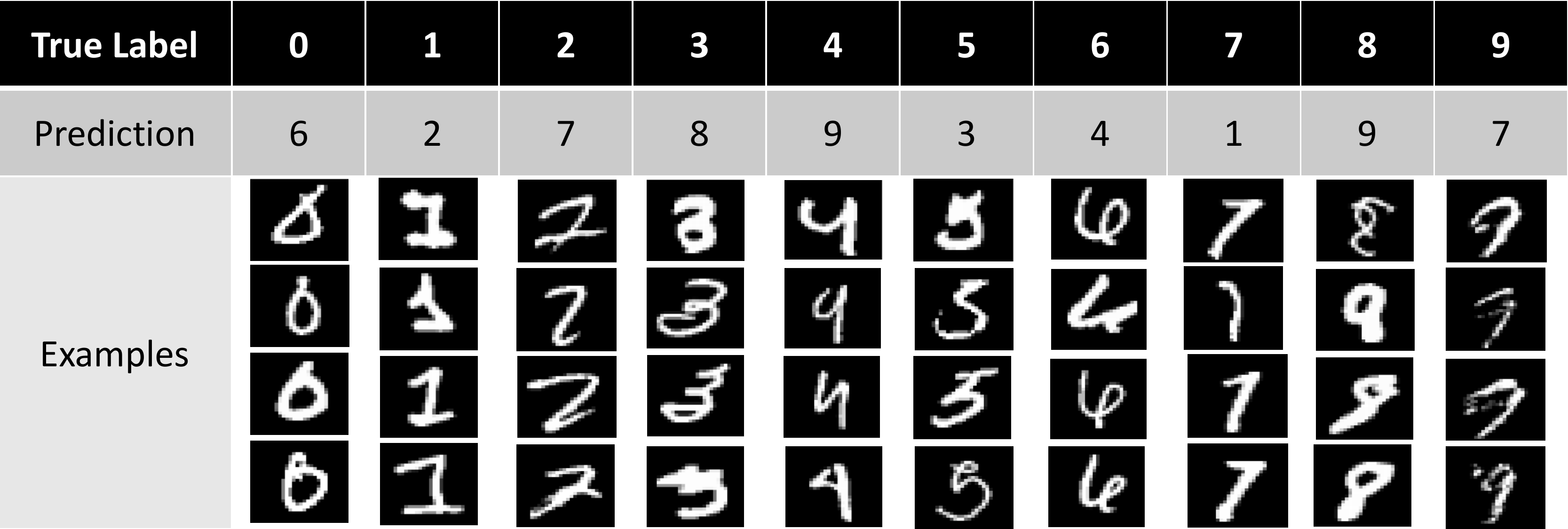}
\caption{Uncertainty Measurement by the Bayesian Neural Network: We show a subset of correctly classified image examples which also predicted a different class in a particular Monte Carlo Sample.}
\label{fig4}
\end{figure}

While the first dot-product in Eq. (3) can be easily performed in a standard RRAM crossbar array \cite{mehonic2019simulation}, the implementation of the second term is shown in Fig. \ref{fig3}. The design consists of two components - (i) \textbf{Stochastic Unit:} An RRAM device undergoes SET-RESET cycling and its resistance is sampled during a subsequent read phase. Fig. \ref{fig2} shows the Normal variation of the sampled conductance value. We utilize a current-to-voltage converter to scale each input to the crossbar array by a factor sampled from the Normal distribution. The amplification factor of this stage is designed such that the maximum read-voltage of the crossbar array is limited to $0.3V$ (to ensure linear RRAM $I-V$ characteristics) \cite{mehonic2019simulation}. Note that these devices are used as binary elements and solely serve the purpose of probability distribution sampling. (ii) \textbf{Deterministic Unit:} The crossbar array stores the parameters of the array $\boldsymbol{\sigma^{\prime}}$. The RRAM devices in the crossbar are programmed to multi-level states which can be achieved by modulating the RESET pulse amplitude, duration or the number of pulses \cite{cheng2017time}. Each column is read sequentially by the array peripherals in order to ensure that independent random samples are being used for each array element. The current flowing through each cross-point device is scaled by the conductance of the device and due to Kirchoff's law, all these currents get summed up along the column, thereby realizing the dot-product kernel. Note that negative $\boldsymbol{\sigma^{\prime}}$ parameters can be also mapped by using two horizontal lines per input (driven by `positive' and `negative' supply voltages). In case a particular parameter is positive (negative), then the corresponding conductance in the `positive' (`negative') line is set in accordance to the parameter. 
The analog current outputs drive interfaced Analog-to-Digital converters (ADCs) to provide output to the fan-out neurons.

A hybrid device-circuit-algorithm co-simulation framework was developed and the design was tested for a standard digit recognition problem on the MNIST dataset \cite{lecun1998gradient}. The neural network used in this work consisted of 2 hidden layers each with 200 neurons. The probability distributions were learnt using the `Bayes by Backprop' algorithm \cite{blundell2015weight} implemented in a PyTorch framework \cite{torch}. The baseline idealized software network was trained with an accuracy of $98.63\%$ over
the training set and $97.51\%$ over the testing set (averaged over
10 sampled networks). Verilog-A RRAM device model \cite{nanoHUB.org19} was used with $45nm$ Predictive Technology Model \cite{ptm} to implement the access transistors. The algorithmic simulator was modified to implement a hardware-aware mapping incorporating probability distribution obtained from device-circuit co-simulation framework (which deviates from the ideal Gaussian distribution), limited bit-precision and programming ranges in the crossbar array elements. We considered 4-bit representation in the RRAM devices in the cross-point array and 3-bit discretization in the Analog-to-Digital converter output (interfaced with the crossbar array). Lower bit precision resulted in degradation of classification accuracy. The RRAM crossbar programmable conductance states were linearly spaced and programmed by modulating the RESET pulse amplitude in the range of $1.4-3.2V$. Recently, more than 6-bit storage has been achieved experimentally in metal-oxide bi-layers \cite{stathopoulos2017multibit}. Note that this work falls into the domain of offline training. The uncertainty measurement functionality is shown in Fig. \ref{fig4} where the depicted images, despite being correctly classified, also generate a different class prediction in a particular Monte Carlo sample. Interestingly, the images also have a degree of similarity to the predicted classes as well. 

While cycle-to-cycle resistance variation of the RRAM elements of the stochastic unit are utilized for computing, other forms of variation are present in the system design. Device-to-device variations of the RRAM elements of the stochastic unit can be present which will result in variation of the parameters $a$ and $b$ of the sampled Normal distribution $N(a,b)$ for each row of the crossbar array. Further, device-to-device variations and write noise can result in variation of the crossbar RRAM resistances. Using our hybrid hardware-algorithm co-simulation framework, the test accuracy of a single sampled network was $98\%$ (over a randomly sampled 100-image subset of the test set). For the variation analysis, the average classification accuracy of 10 independent Monte Carlo runs, assuming $10\%$ Gaussian noise on each of the above-mentioned parameters (each run consisting of a single sampled network over the same 100 image test-set), was evaluated to be $93\%$. Note that this accuracy degradation can be potentially reduced by exploring hardware-in-the-loop training techniques. 

Our proposal of partitioning the computation as described in Eq. (3) and hybrid system design with stochastic and deterministic units allows us to perform the costly dot-product operations (one of the main computationally expensive operations in a neural network hardware) in the memory array itself, enabling us to address the issues of von-Neumann bottleneck. From a system design perspective, the energy consumption is dominated by the Gaussian random number generation process since it involves RRAM programming while the remaining components are primarily RRAM read operations and peripheral contributions. Hence, to substantiate our benefits in contrast to CMOS hardware, we first focus on the Gaussian random number generation hardware complexity. CMOS based implementations of Gaussian random number generators rely on a significant number of linear feedback circuits which are extremely hardware expensive. For instance, a recent work for a CMOS based $64$-parallel Gaussian RNG reports $1780$ registers and $528.69mW$ power consumption \cite{cai2018vibnn}. In stark contrast, our implementation uses simply a single RRAM device and a current-to-voltage converter to implement the probability distribution sampling. The total power consumption for our RRAM based Gaussian random number generation is $32.54mW$ ($VI$ power consumption for the stochastic unit circuit during a complete cycle RRAM operation to generate a sample from a Gaussian distribution) for a similar 64 parallel generation task. For system level energy consumption evaluation, we considered 8-bit resolution in DAC/ADC designs to achieve iso-bit length comparison with Ref. \cite{cai2018vibnn}. Typical peripheral energy consumption metrics for DAC, ADC and Sample and Hold circuits reported in Refs. \cite{ankit2019puma,shafiee2016isaac} were included. The crossbar read-latency per column was considered to be $10ns$. The overall energy efficiency of the system was evaluated to be $106567.2$ $Images/J$, which is $2\times$ higher than the baseline CMOS implementation \cite{cai2018vibnn}.  

It is worth noting here that device level non-idealities in other technologies can be harnessed as well for such probabilistic AI hardware. 
This work is based on Ref. \cite{yang2019all} which explored the design for spintronic technologies. This proposal explores the design for RRAM devices which requires a significant rethinking due to different intrinsic device physics and operating voltage/current characteristics and conditions. The key distinguishing factor of the RRAM based design over the spintronic implementation lies in the fact that the Gaussian random number sampling (the most hardware expensive component of the overall system) can be directly emulated in a single device. Further, RRAM based crossbar arrays are much more mature in terms of scalable hardware implementation than spin-based neuromorphic arrays, thereby being amenable for near-term demonstration.

\section{Summary}
In conclusion, we have provided a proposal for an RRAM based ``In-Memory" computing primitive for Bayesian neural hardware that utilizes simple device-circuit primitives and leverages RRAM stochasticity as a computing resource, instead of viewing it as a disadvantage. Such hardware-software co-design of Bayesian neural network models can potentially lead to real-time decision making in autonomous agents in the presence of uncertainties.  


\end{document}